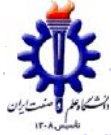

# The Effect of Wedge Tip Angles on Stress Intensity Factors in the Contact Problem between Tilted Wedge and a Half Plane with an Edge Crack Using Digital Image Correlation

[1]S. Khaleghian[*], [2]A. Emami, [3]M. Yadegari and [4]N. Soltani

[1,3] *Master Student,* [2] *Bachelorer Student,* [4] *Professor, Department of Mechanical Engineering,
University of Tehran, Tehran, Iran*
[*](Tel: (98) 912 7127609, e-mail: sm.khaleghian@yahoo.com)

**Abstract**

The first and second mode stress intensity factors (SIFs) of a contact problem between a half-plane with an edge crack and an asymmetric tilted wedge were obtained using experimental method of Digital Image Correlation (DIC). In this technique, displacement and strain fields can be measured using two digital images of the same sample at different stages of loading. However, several images were taken consequently in each stage of this experiment to avoid the noise effect. A pair of images of each stage was compared to each other. Then, the correlation coefficients between them were studied using a computer code. The pairs with the correlation coefficient higher than 0.8 were selected as the acceptable match for displacement measurements near the crack tip. Subsequently, the SIFs of specimens were calculated using displacement fields obtained from DIC method. The effect of wedge tip's angle on their SIFs was also studied. Moreover, the results of DIC method were compared with the results of photoelasticity method and a close agreement between them was observed.

## 1. Introduction

Digital Image Correlation (DIC) has become very popular among experimental-optical analysis methods. Its simple set-up comparing to other methods such as shearography, moiré, caustic method etc., and its vast use in stress and strain analysis has made it a practical technique. It is also one of the most reliable and precise methods in experimental analyzing. This method was first used at the University of South Carolina for measuring the whole displacement field by Peters and Ranson [1]. In a few years, Sutton et al. [2] improved this method and showed its precision in obtaining strain field. They used Newton-Raphson method for optimizing the process [3]. Venroux [4] optimized the basic algorithm for displacement and displacement gradient measurements. Vendroux and Knauss [5] used this method for submicron measurement of strain. Since then, many researchers have been working in this field and developed the method [6-8]. Jin and Bruck [9] extended a new point-wise DIC method, which could measure the discontinuous displacements. Réthoré et al [10, 11] developed another DIC method, for crack analysis. Sutton et al. [12] used 3D DIC set-up to demonstrate the effect of out-of-plane displacement on 2D DIC measurements.

Several researchers also tried to improve the precision of DIC methods. Some of them worked on software and programs for correlation process of images. For instance, Poissant and Barthelat [13] developed in-house MATLAB codes. Others worked on random patterns created on the surface of specimens. These patterns are mostly created by spraying white and black color in order to produce random speckles. Hua et al. [14] studied the characteristics of the optimum random pattern for DIC specimens.

Since commencement of developing this method, many researchers have been engaged in this area. Due to the sensitivity and precision of the method, it has been used in different scopes [15]. Determining the stress intensity factors is one of the most important applications of this method. It could be determined by using displacement field and finite element or finite deference method [16]. Since then many researchers have worked on this area [17-20].

In this investigation, DIC method was used to determine the SIFs in a contact problem between half planes with 60° and 90° edge crack and two different asymmetric tilted wedges. This problem was developed to simulate the screw and rivet joints in wings and fuselage aircrafts which is a major concern with the aging and fretting wear in aerospace industry. Hence, the effect of wedge tip's angles and edge crack's angles on the SIFs of half plane under three different contact loadings was studied. The results were also compared with the photoelasticity technique and good agreement between SIFs, obtained through these two different experimental methods, was observed.

## 2. Experimental details

### 2.1 Specimens

Two plates with 240-mm length, 60-mm width, and 4-mm thickness were used to model the half planes. The dimensions were designated using the FEM simulations having no effect on the calculated SIFs. The plates were made of polycarbonate provided by Bayer Co. of Germany. This polycarbonate has an elastic modulus of 2400 MPa, Poisson's ratio of 0.38, and friction coefficient of 0.4. The edge crack of plates was inclined at 60° and 90° with respect to the horizontal axis. The cracks length was 5 mm by 0.2 mm width which were machined using a 0.2-mm blade into the plate.

Two asymmetric tilted wedges made of steel CK45 were used in the experiments to contact with plates. Both wedges had total tilt angle of 165° at their apexes but with different side angles with respect to the vertical axis ($\theta_1$ and $\theta_2$ as shown in Fig. 1). The schematic of asymmetric tilted wedges is shown in Figure 1. The magnitudes of $\theta_1$ and $\theta_2$ for wedge labeled 'W1' are 87 and 78 degree respectively, and for wedge labeled 'W2' are 84 and 81 degree.

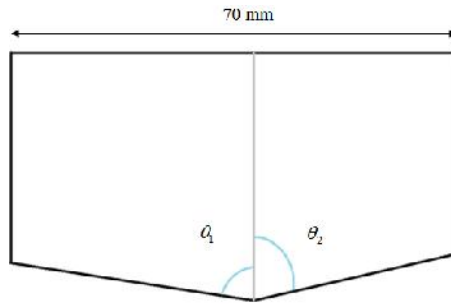

Figure 1: Schematic of asymmetric wedge geometry

The difference in wedges' side angles effects on the distribution of contact force in horizontal and vertical directions. The horizontal force applied by wedges increases as the difference between tip's side angles of tilted wedges increases. The distance between contact point and crack opening was 4mm in all cases and the applied load for each specimen was varied as 188, 255, and 305N. Schematic of the model used for this contact problem is shown in Figure 2.

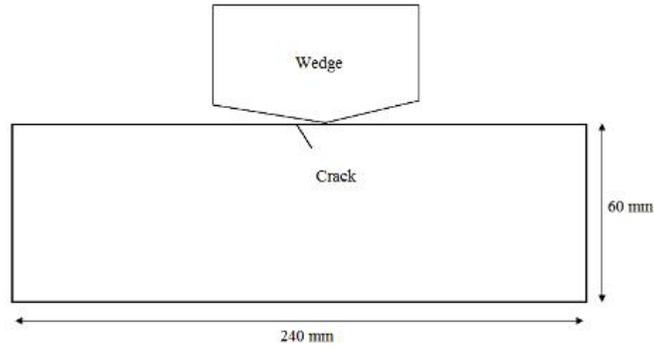

Figure 2: Illustration of the model used for contact problem

**2.2 DIC Method**

The experiment was performed using the set-up shown in Figure 3 with a load cell of 200Kg.F capacity. An art-cam 320p CCD camera equipped with a Fujian lens was used to take the images. The images were captured with the spatial resolution of 2288×1700 pixels. A standard white light source was also used to illuminate the surface. The set-up was settled on a table with vibration isolation scheme in order to eliminate the noises.

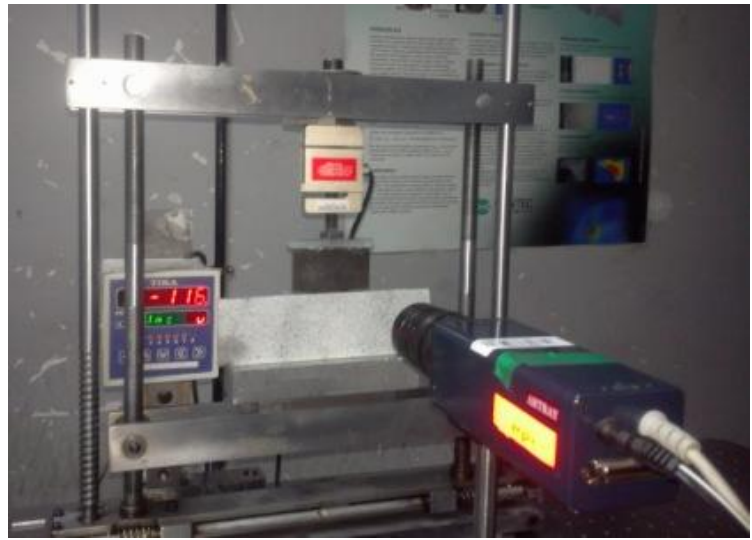

Figure 3: Set-up of DIC experiment

All the specimens had been prepared before the experimental procedures and their surfaces had been sprayed by white and black color to create random speckle patterns on them. Each specimen located in the set-up and images were capture before and after the loading. In each experiment, five images, instead of one, were captured in each step, before and after loading, in order to reduce the effect of the light's noise on the images. Then, the final image (the average of five images) was recorded in 8-bit gray scale. A pair of images for each experiment was considered and compared with each other in order to study the correlation between them. The correlation program was used to analyze the images acquired at increasing loading of each specimen. The correlation equation used in this work can be defined as:

$$C(x,y,u,v) = \frac{\sum_{i=-n}^{i=n}\sum_{j=-n}^{j=n}\left(I_R(x_p+i, y_p+j) - I_D(x_p+i+u_p, y_p+j+v_p)\right)^2}{\sum_{i=-n}^{i=n}\sum_{j=-n}^{j=n}\left(I_R(x_p+i, y_p+j)\right)^2} \qquad (1)$$

where IR and ID represent the light intensity before and after the loading respectively and C is the correlation coefficient while 1-C is the representation of correlation quality. Therefore, the correlation quality would be from 0 to 1, in which 1 indicates the perfect match.

The images with correlation quality higher than 0.8 were considered as the most reliable matches and the displacement and strain fields were obtained from them. Coordinates of 60 pixels on the region near the crack tip were extracted by means of a code written in MATLAB software which provides the real scale of images using actual length of the crack. Then, the SIFs were found using the displacement fields in the following equations [21] which were solved by the algorithm of least-squares estimation:

$$U = \frac{1}{2\mu}\sqrt{\frac{r}{2\pi}}\left(K_I \cos(\frac{\theta}{2})\left[k-1+2\sin^2(\frac{\theta}{2})\right] + K_{II}\sin(\frac{\theta}{2})\left[k+1+2\cos^2(\frac{\theta}{2})\right]\right) + \alpha r\cos_\theta + U_o$$
$$V = \frac{1}{2\mu}\sqrt{\frac{r}{2\pi}}\left(K_I \sin(\frac{\theta}{2})\left[k+1-2\cos^2(\frac{\theta}{2})\right] - K_{II}\cos(\frac{\theta}{2})\left[k-1-2\sin^2(\frac{\theta}{2})\right]\right) + \alpha r\sin_\theta + V_o$$
(2)

where U and V are the displacement components in x and y directions respectively. μ gives the shear modulus which is E/2(1+ν) where E is elastic modulus and ν is the Poisson's ratio. k is (3-ν)/(1+ν) for plane stress and 3-4ν for plane strain. r and θ express the polar coordinates around the crack tip, and The constants $U_0$, $V_0$, and α are associated with rigid body displacements and rotations.

**2.3 Photoelasticity Method**

Photoelastic data were obtained using the specimens in a polariscope which consisted of a monochromatic red light source, a polarizer, quarter-wave plates, and an analyzer. A 7-mega pixel professional color camera was also used to capture images of photoelastic fringes formed in the plates under contact loading applied by asymmetrical wedges. Since the widths of isochromatic fringes in the photoelastic images were non-zero and therefore the exact positions of dark fringe paths were not discernible, all the captured images were processed and sharpened by an image processing computer code. This code is able to detect the exact locations of pixels on the isochromatic fringes where the light intensity was extremum. A sample of image before and after processing is shown in Figures 4.a and 4.b respectively. This image is for the half plane with 60° edge crack and W2 under 305N loading.

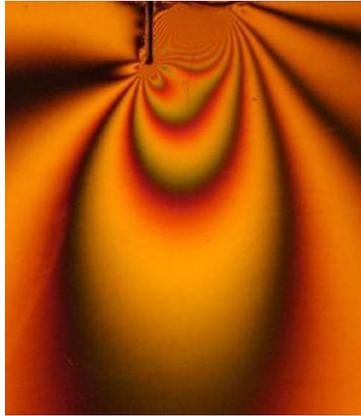
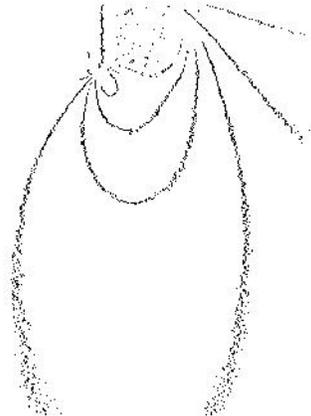

Figure 4.a: Sample photo before image processing    Figure 4.b: Sample photo after image processing

After image processing, each image was loaded by a code written in MATLAB software which detected the actual size of the model in the image and determined the actual polar coordinates of any selected pixel in it by locating the origin of coordinate system on the crack tip and r-axis in direction of crack length. Then, 20 pixels of the isochromatic fringes were selected in each image and their coordinates were extracted to minimize the errors in calculation of the SIFs in the conducted experiments [22].

Linear elastic fracture mechanics (LEFM) gives the following equations for calculating the stress field around the crack tip in the polar coordinate system [23]:

$$\sigma_r = \frac{K_I}{4\sqrt{2\pi r}}(5\cos(\frac{\theta}{2}) - \cos(\frac{3\theta}{2})) + \frac{K_{II}}{4\sqrt{2\pi r}}(-5\sin(\frac{\theta}{2}) + 3\sin(\frac{3\theta}{2}))$$

$$\sigma_\theta = \frac{K_I}{4\sqrt{2\pi r}}(3\cos(\frac{\theta}{2}) + \cos(\frac{3\theta}{2})) + \frac{K_{II}}{4\sqrt{2\pi r}}(-3\sin(\frac{\theta}{2}) - 3\sin(\frac{3\theta}{2})) \quad (3)$$

$$\tau_{r\theta} = \frac{K_I}{4\sqrt{2\pi r}}(\sin(\frac{\theta}{2}) + \sin(\frac{3\theta}{2})) + \frac{K_{II}}{4\sqrt{2\pi r}}(\cos(\frac{\theta}{2}) + 3\cos(\frac{3\theta}{2}))$$

where $r$ and $\theta$ are coordinates of the point in vicinity of crack tip, $K_I$ and $K_{II}$ are SIFs of mode I and II. Besides, the stress-optic law for a plane stress case is defined as [24]:

$$\tau_{max} = \frac{Nf_\sigma}{2h} \quad (4)$$

where $\tau_{max}$ is the maximum shear stress, N is the relative retardation in terms of a complete cycle of retardation (isochromatic fringe order), and $f_\sigma$ is the material fringe value which was acquired from the calibration test. The maximum in plane shear stress can be defined as:

$$(\tau_{max})^2 = (\frac{\sigma_r - \sigma_\theta}{2})^2 + (\tau_{r\theta})^2 \quad (5)$$

Combining Equations (4) and (5) leads to:

$$(\frac{Nf_\sigma}{h})^2 = (\sigma_r - \sigma_\theta)^2 + 4(\tau_{r\theta})^2 \quad (6)$$

The over-determined least square estimation method was employed to calculate the SIFs using the combination of Equation (1) and Equation (3) with polar coordinates and fringe numbers (r, $\theta$, and N) of selected pixels.

## 3. Results and Discussion

As a sample case, the contour maps and 3D-graphs of the displacement fields around the crack tip obtained by digital image correlation for half plane with 90° edge crack are presented in this section. Figure 5 and Figuare 6 show the images related to contact of the half plane with W1 and W1 respectively. The vertical force applied on the specimen was 255N. Note that in the 3D-graphs, X1 and X2 represent horizontal and vertical axis respectively.

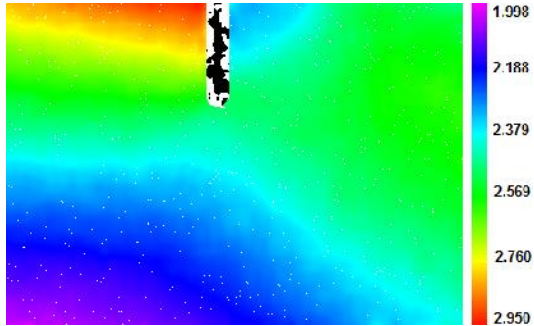 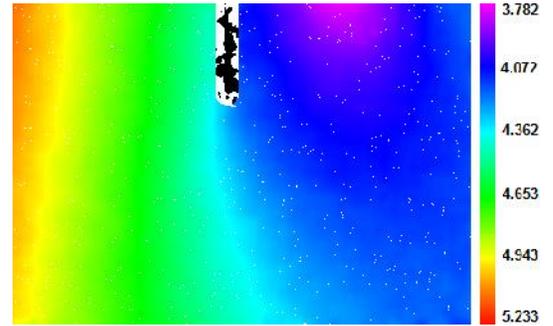

Figure 5.a: Contour map of horizontal displacement field around the crack tip for the contact between the half plane and W1

Figure 5.b: Contour map of vertical displacement field around the crack tip for the contact between the half plane and W1

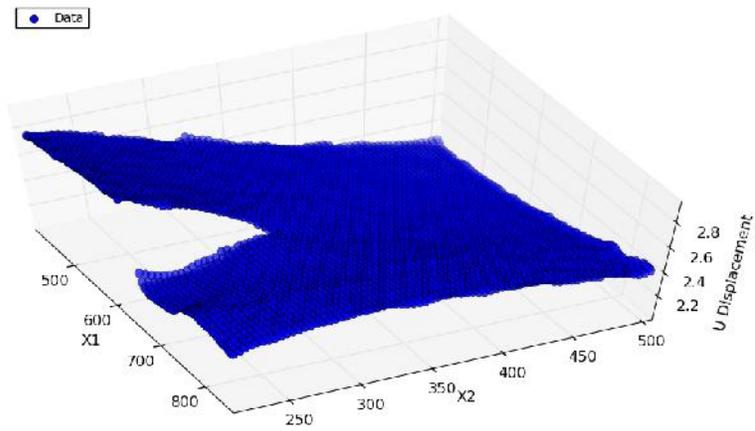

Figure 5.c: 3D-graph of horizontal displacement field around the crack tip for the contact between the half plane and W1

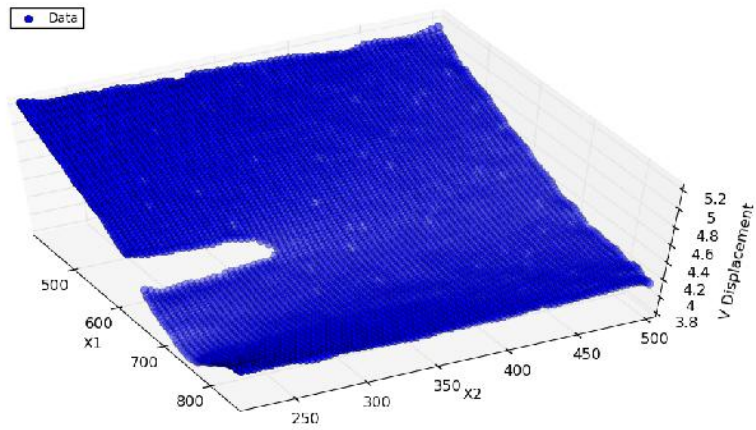

Figure 5.d: 3D-graph of vertical displacement field around the crack tip for the contact between the half plane and W1

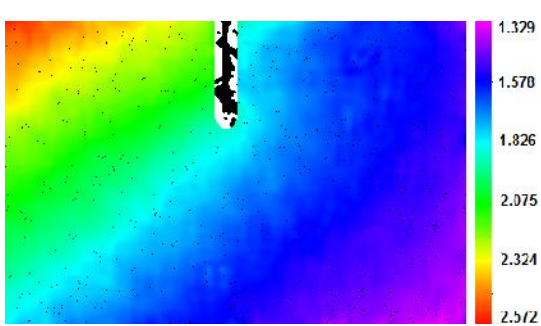 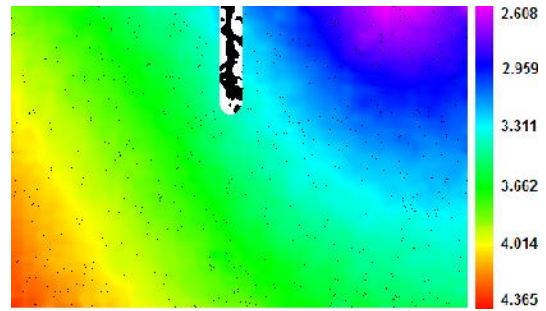

Figure 6.a: Contour map of horizontal displacement field around the crack tip for the contact between the half plane and W2

Figure 6.b: Contour map of vertical displacement field around the crack tip for the contact between the half plane and W2

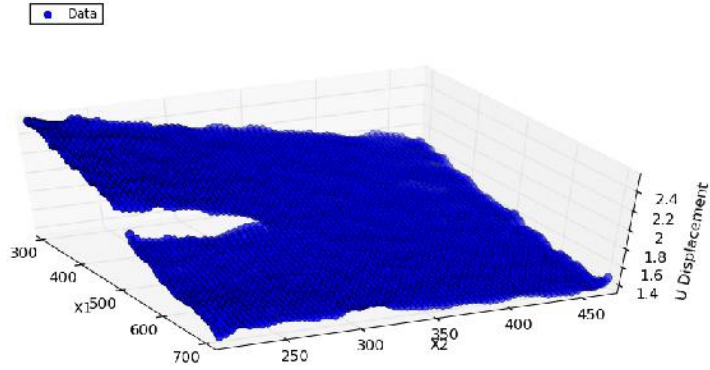

Figure 6.c: 3D-graph of horizontal displacement field around the crack tip for the contact between the half plane and W2

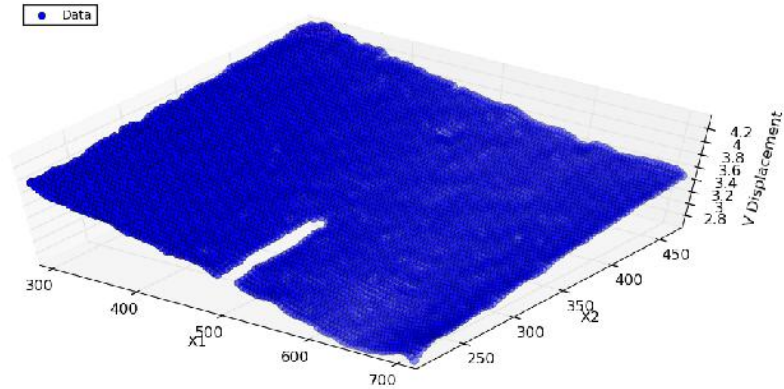

Figure 6.d: 3D-graph of vertical displacement field around the crack tip for the contact between the half plane and W2

All the SIFs found through least squares estimate of experimental data along with the relative difference of SIFs obtained from DIC method and photoelasticy method are presented in Table 1 and Table 2. Note that in this contact problem, the half plane with 90° edge crack was in the mixed mode of fracture and had both in-plane SIFs of $K_I$ and $K_{II}$; however, the other half plane, which had 60° edge crack, were in the second mode of fracture for all the cases and the values of $K_I$ for them were all zero.

Table 1: Half plane with 60° edge crack

| Wedge | Force (N) | $K_{II}$ | | |
|---|---|---|---|---|
| | | DIC Method (KPa× $\sqrt{m}$) | Photoelasticity Method (KPa× $\sqrt{m}$) | Relative Difference (%) |
| W1 | 188 | 150.728 | 148.3105 | 1.63 |
| | 255 | 151.3007 | 148.5525 | 1.85 |
| | 305 | 166.6248 | 162.9901 | 2.23 |
| W2 | 188 | 103.0732 | 101.2606 | 1.79 |
| | 255 | 111.2559 | 108.7439 | 2.31 |
| | 305 | 125.2182 | 122.0926 | 2.56 |

Table 2: Half plane with 90° edge crack

| Wedge | Force (N) | $K_I$ | | | $K_{II}$ | | |
|---|---|---|---|---|---|---|---|
| | | DIC Method (KPa×√m) | Photoelasticity Method (KPa×√m) | Relative Difference (%) | DIC Method (KPa×√m) | Photoelasticity Method (KPa×√m) | Relative Difference (%) |
| W1 | 188 | 482.644 | 475.6519 | 1.47 | 249.5334 | 246.2825 | 1.32 |
| | 255 | 602.7072 | 592.6907 | 1.69 | 315.9526 | 311.0382 | 1.58 |
| | 305 | 838.1133 | 819.9113 | 2.22 | 441.743 | 434.2308 | 1.73 |
| W2 | 188 | 305.3787 | 301.8172 | 1.18 | 184.5025 | 182.7481 | 0.96 |
| | 255 | 420.2001 | 413.9903 | 1.5 | 266.7465 | 263.5835 | 1.2 |
| | 305 | 549.5321 | 538.0185 | 2.14 | 375.7908 | 368.8925 | 1.87 |

The relative difference between results of DIC method and photoelasticy method is less than 3% in all cases. As it is shown in Figure 7, this relative difference was raised by increase of vertical force. Therefore, when the vertical force has small magnitude and photoelastic method could not give accurate results due to the lack of detectable isochromatic fringes near the crack tip, the DIC method could be used instead.

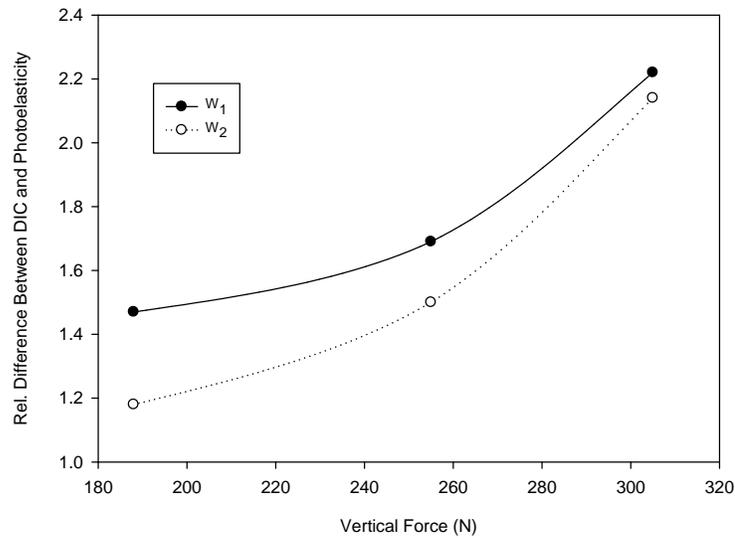

Figure 7: The relation between vertical force and relative difference between $K_I$ obtained from DIC method and photoelasticy method for half plane with 90° edge crack

The relation between vertical force and relative difference between KII obtained from DIC method and photoelasticy method for both half planes is shown in Figures 8.a and 8.b.

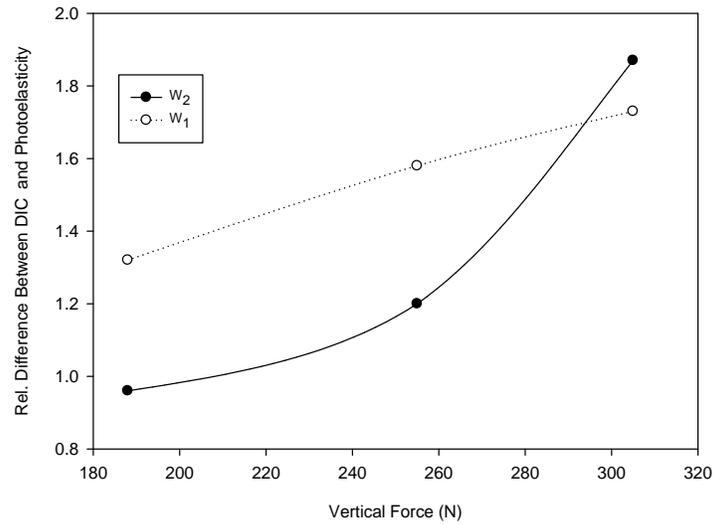

Figure 8.a: The relation between vertical force and relative difference between $K_{II}$ obtained from DIC method and photoelasticy method for half plane with 90° edge crack

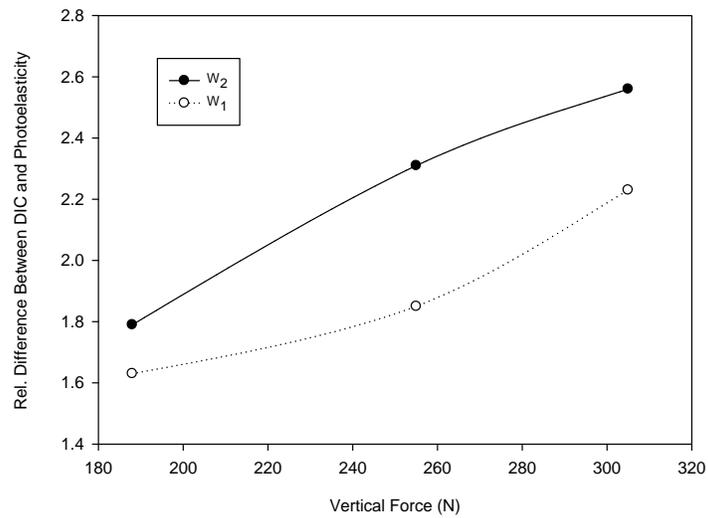

Figure 8.b: The relation between vertical force and relative difference between $K_{II}$ obtained from DIC method and photoelasticy method for half plane with 60° edge crack

As Figures 8.a and 8.b indicate, the relative difference between results of DIC method and photoelasticy method for half plane with 90° edge crack was not as much as half plane with 60° edge crack. Thus, the results of half plane with 90° edge crack are more reliable than the other one.

The relation between the second mode SIF ($K_{II}$) of half plane with 60° edge crack and the vertical force applied on four wedges with different tip's side angles is shown in Figure 9. The SIF values obtained from DIC method corresponded well to the values gained through photoelasticity experiment; however, DIC method gave higher SIFs values for all the cases.

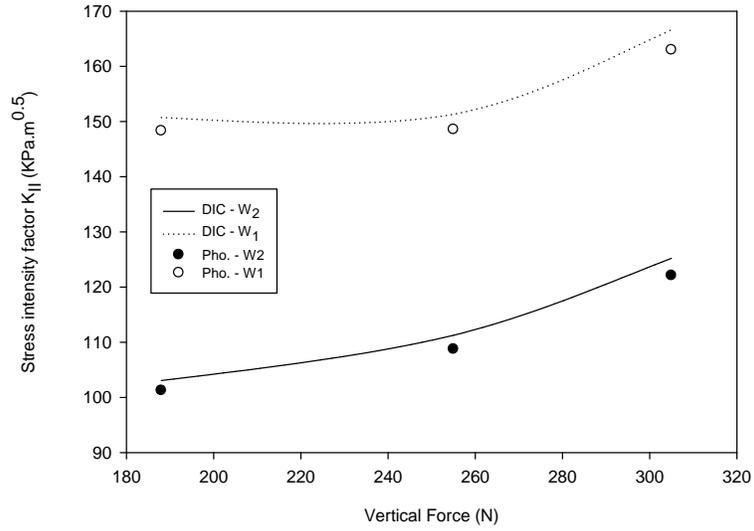

Figure 9: The relation between the second mode SIF ($K_{II}$) of half plane with 60° edge crack and the vertical force applied on four wedges with different tip's side angles

As it was predicted, the increase of vertical force resulted in higher values of $K_{II}$ due to the increment of in-plane displacement tendency. Moreover, the horizontal force had the same effect on $K_{II}$ values. As it was mention before, the horizontal force applied by wedges has direct relation with the difference between tip's side angles of tilted wedges; therefore, the horizontal force of W1 and W2 as well as their corresponding $K_{II}$ values decreased respectively.

The effect of vertical force on $K_I$ and $K_{II}$ in half plane with 90° edge crack for four wedges used in the experiments is shown in Figure 10 and Figure 11 respectively. Note that in all cases the value of $K_I$ is higher than $K_{II}$ which indicates the half plane was more likely to break under the first mode of fracture.

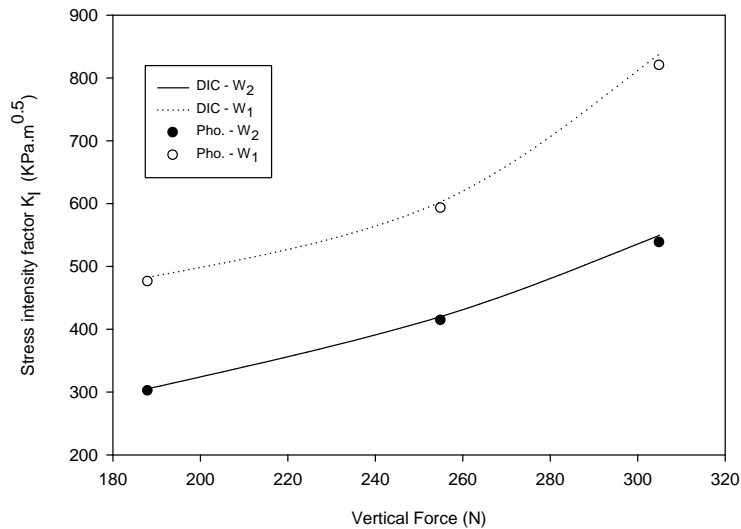

Figure 10: The effect of vertical force on SIF ($K_I$) of half plane with 90° edge crack for four wedges used in the experiments

As it is displayed in Figure 10, the increase of vertical force affects the opening mode of crack and raises the value of $K_I$. In contrast to 60° edge crack, the 90° edge crack tended to rupture under the horizontal force applied by the wedges used in the experiments; as a consequent, $K_I$ had the higher value while W1 was used in the experiment.

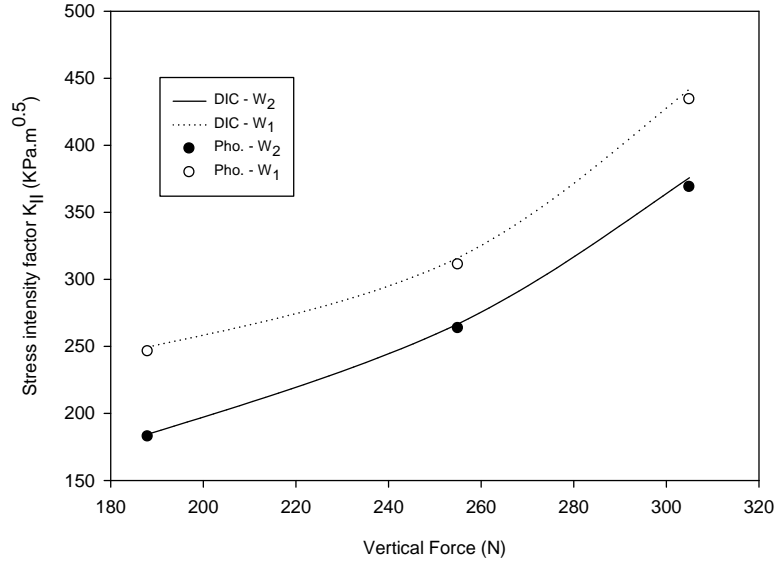

Figure 11: The effect of vertical force on SIF ($K_{II}$) of half plane with 90° edge crack for four wedges used in the experiments

As it is presented in Figure 11, the increase of vertical force leads to more in-plane displacement and consequently higher value of $K_{II}$. Likewise, the increase of horizontal force causes higher in-plane shear SIF ($K_{II}$); therefore, while the half plane was in contact with wedge W1, $K_{II}$ had higher value.

## 5. Conclusion

DIC method was utilized for evaluation of SIFs in the contact problem between a half plane with an edge crack and an asymmetric tilted wedge. The effect of wedge tip angles on SIF values was studied for two different crack angles (60° and 90°) under three different force magnitudes. Then, the results of DIC method were compared with the photoelastic results. The relative difference between SIFs' values obtained from DIC and photoelasticity was less than 3% for all the cases. In general, the relative difference between the results of these two methods for half plane with 90° edge crack was lesser than that of plane with 60°crack. This relative difference increased as more vertical force applied on the specimens. Therefore, DIC method could be a good alternative for photoelasticity when the magnitude of applied force and consequently the number of isochromatic fringes decreases to such a low number that the necessary date for accurate results could not be extracted from photoelastic images.

     A comparison between two half planes indicates that in this type of loading, the half plane with 60° edge crack was in the second mode of fracture; conversely, the half plane with 90° edge crack was in the mixed mode of fracture and its first mode SIF had greater value than the second one. However, the increase of vertical force had the same effect on SIFs of both half planes and raised their values, but never led the 60° edge crack to go to the first mode of fracture and its $K_I$ value remained zero. Additionally, the horizontal force was raised by increasing the difference between wedge side angles. The increase of horizontal force similar to the increase of vertical force had the same effect on SIFs as vertical force did. Thus, it can be concluded that as the asymmetry of in-contact wedge increases, the half plane will break under less force magnitude.